\documentclass[aps,prl,twocolumn,showpacs,superscriptaddress,groupedaddress]{revtex4}

\usepackage{array,multirow,graphicx}  
\usepackage{dcolumn} 
\usepackage{bm}
\usepackage{amssymb}
\usepackage[utf8]{inputenc}
\usepackage{pbox}
\usepackage{float}
\usepackage[caption = false]{subfig}
\usepackage{color} 

\begin{document}

\title{Secondary Electron Emission from Surfaces with Small Structure}

\author{A. R. Dzhanoev}
\affiliation{Universit\"at Potsdam, Karl-Liebknecht-Str. 24/25 Building 28, 14476 Potsdam-Golm, Germany}

\author{F. Spahn}
\affiliation{Universit\"at Potsdam, Karl-Liebknecht-Str. 24/25 Building 28, 14476 Potsdam-Golm, Germany}
 
\author{V. Yaroshenko}
\affiliation{GFZ, German Research Centre for Geosciences, Telegrafenberg, 14473 Potsdam, Germany}

 \author{H. L\"uhr}
 \affiliation{GFZ, German Research Centre for Geosciences, Telegrafenberg, 14473 Potsdam, Germany}
 
 \author{J. Schmidt} 
\affiliation{University of Oulu, Astronomy and Space Physics, PL 3000, Oulu, Finland}

\date{\today; http://journals.aps.org/prb/abstract/10.1103/PhysRevB.92.125430}

\begin{abstract}
It is found that for objects possessing small surface structures with differing radii of curvature the secondary electron emission (SEE) yield may be significantly higher than for objects with smooth surfaces of the same material. The effect is highly pronounced for surface structures of nanometer scale, often providing a more than $100 \%$ increase of the SEE yield. The results also show that the SEE yield from surfaces  with structure does not show an universal dependence on the energy of the primary, incident electrons as it is found for flat surfaces in experiments. We derive conditions for the applicability of the conventional formulation of SEE using the simplifying assumption of universal dependence. Our analysis provides a basis for studying low-energy electron emission from nano structured surfaces under a penetrating electron beam important in many technological applications.
\end{abstract}

\pacs{68.37.-d, 79.20.Hx, 94.05.-a}
\maketitle
\section{Introduction}
Secondary electron emission from solids by electron bombardment has been subject of experimental and theoretical studies for many decades \cite{Salow,Baroody, Jonker1952, Sternglass1954, Kollath, Sternglass1957, Hachenberg, HallBeeman, DraineSalpeter1979, Seiler1983, Chowetal1993,Millet,Scholtz,Young,Jablonski,Fitting, Wintucky, Joy, Nishimura,Ziemann,Shih1997, Caron, Reimer,Richterova}, with a wide variety of applications including voltage contrast in scanning electron microscopy, micro channel plates, plasma display panels, electron beam inspection tools. In space, attention has focused on SEE from spacecraft surfaces or small dust particles caused by auroral electrons or hot electrons in planetary magnetospheres \cite{Horanyi, Spahnetal2006,Kempf,JSchmidt2008}. 
To date, theoretical studies of SEE usually involve a slab model which often gives a reasonable estimate of the SEE yield (but see also \cite{DraineSalpeter1979, Chowetal1993, Millet} and discussion below). In this paper, we show that the interplay between the penetration depth of primary electrons, the escape depth of secondary electrons and the size of surface structure (surface curvature) can be the dominant mechanism for the SEE from small objects. Moreover the SEE from configurations involving nano structures are of fundamental importance for basic science as well as for 
applications \cite{DraineSalpeter1979, Seiler1983, Chowetal1993} ranging from astrophysics to technological processes. The detailed understanding and proper interpretation of the SEE yield from either nano-scaled structures on surfaces or nano-sized objects is strongly desired because such knowledge is, for instance, important for scanning electron microscope imaging of small objects and any charging processes where secondary electron currents are involved.
 
The total electron yield is often written as a sum $\sigma=r+\eta+\delta$ of elastically ($r$) and inelastically ($\eta$) backscattered electrons (BSE) as well as true secondary electrons (SE) ($\delta$). For incident electrons with energies in the range where the SEE dominates (typically above $100\text{ eV}$) elastically and inelastically backscattered primaries ($R=r+\eta$) constitute only a small fraction of the total yield \cite{Reimer}. In the literature, it is common to use the simplifying assumption of SEE from a large, planar sample \cite{Baroody, Jonker1952, Sternglass1954, Kollath, Sternglass1957, Hachenberg, HallBeeman, DraineSalpeter1979, Chowetal1993}. It has been suggested that the energy dependence of the SEE yield can be described by the Sternglass universal curve \cite{Sternglass1957}, when the yield is normalized by the maximum yield and the primary electron energy by the energy where the yield is maximized \cite{Baroody}. However a series of measurements of SEE covering a wide range of primary energies \cite{Salow,Young} shows that the theory of Sternglass fails to fit the experimental data at high primary electron  energy. The empirical formula developed by Draine and Salpeter \cite{DraineSalpeter1979} generally shows a better agreement with experiments but overestimates the data. Later Chow et al. \cite{Chowetal1993} have modified the yield equation by Jonker \cite{Jonker1952} and derived the yield for secondary emission from a spherical dust grain immersed in a plasma environment. The influence of porosity on electron-induced SEE was considered by Millet and Lafon \cite{Millet}. There are also some numerical models of SEE from dust grains \cite{Richterova}. However all these approaches assume a smooth surface of the object. The first attempt to include surface structures have been made by Nishimura et al. \cite{Nishimura}. Their Monte Carlo simulations have shown that neglecting the surface roughness may considerably underestimate the magnitude of the secondary electron yield. These results are in a good agreement with the characteristics of low secondary and reflected primary electron emissions from textured surfaces measured by Wintucky et al. \cite{Wintucky}.

\section{Model}
In this paper, we study the effect of small spatial surface structures of three-dimensional samples on the SEE efficiency. We suggest an analytical expression for the SEE yield accounting for the effect of surface curvature in an approach generalizing existing elementary SEE models. We show that the presence of small structures on a sample surface destroys the universal dependence of yield on energy. Moreover, we find a significant growth of the electron yield. 

\begin{figure}[h!]
{\bf(a)} \hfill \vspace{1mm}

{\includegraphics[scale=0.3]{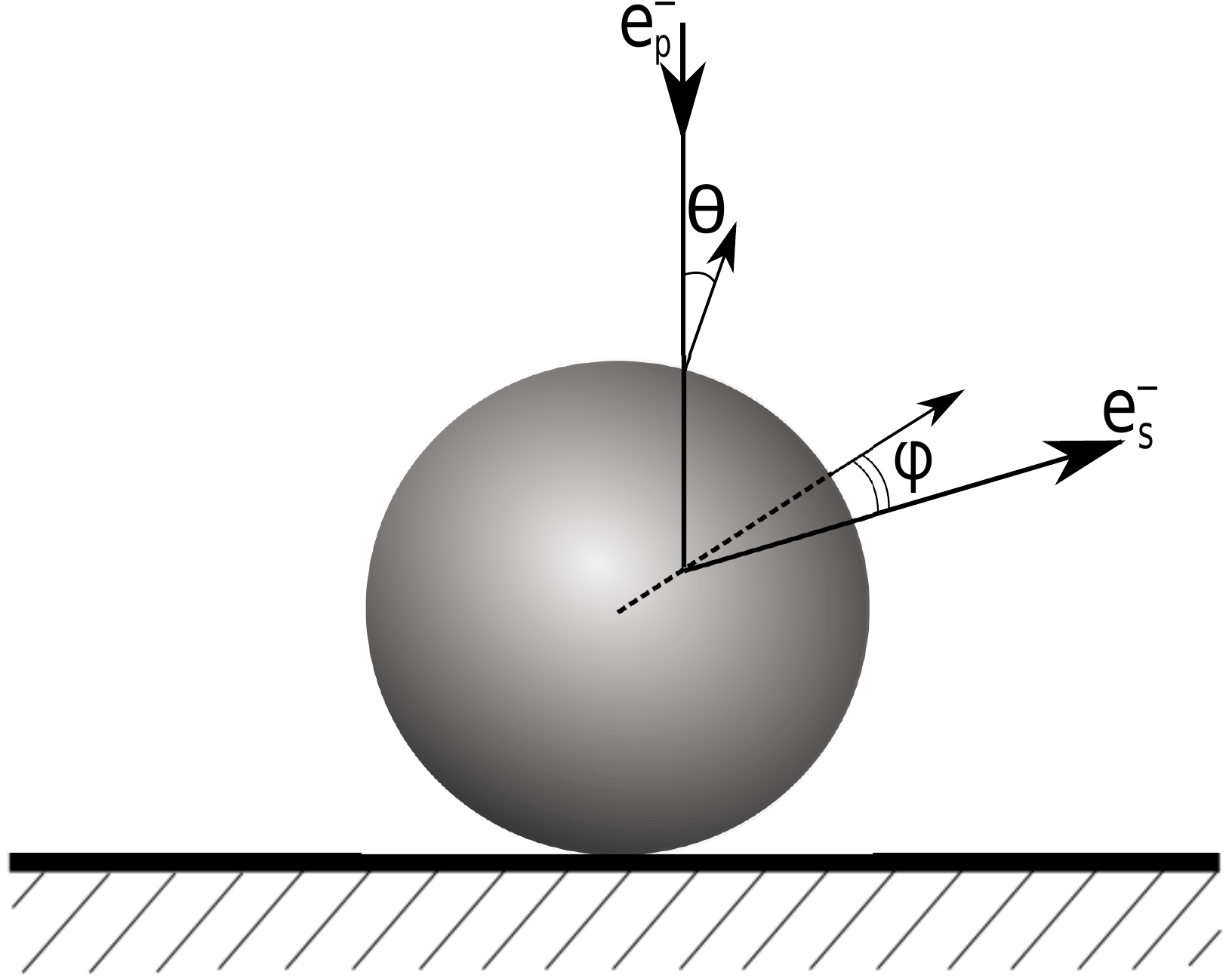}}
 
{\bf(b)} \hfill \vspace{3mm}

{\includegraphics[scale=0.3]{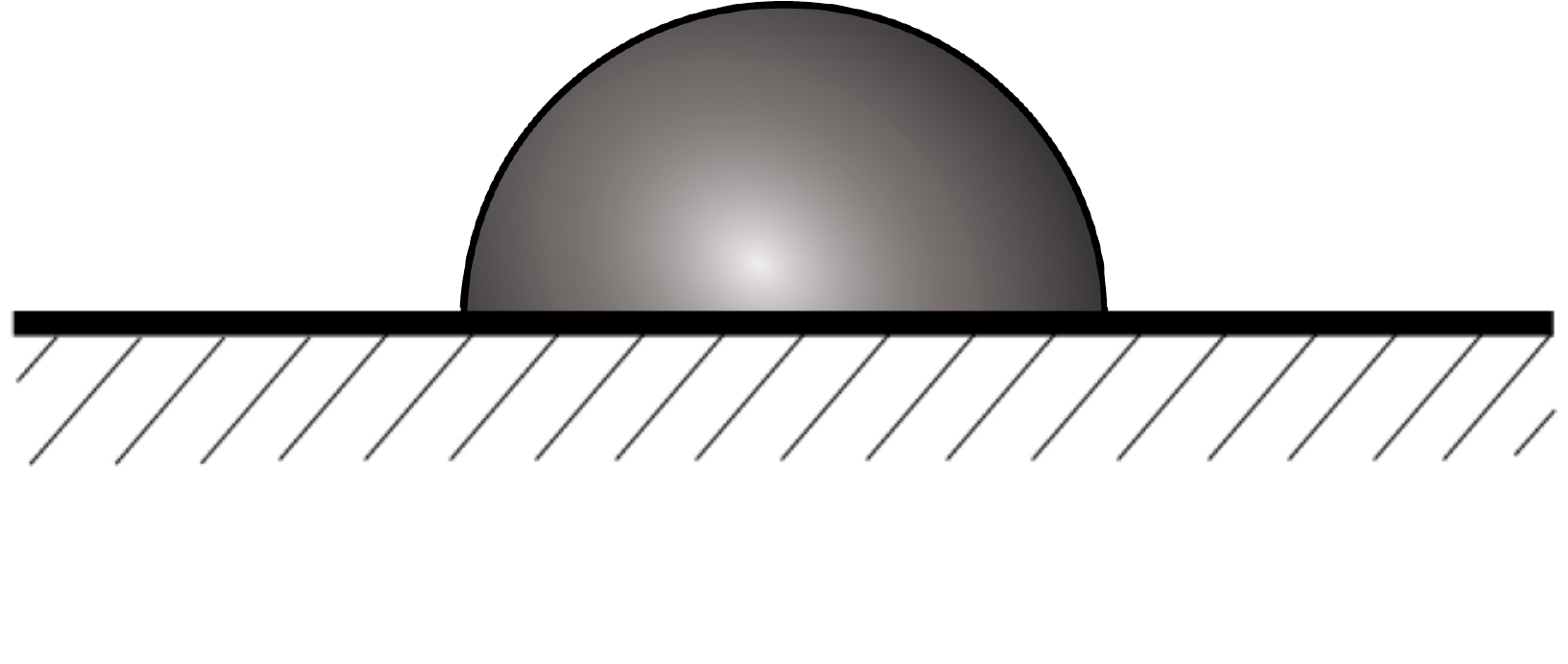}}

{\bf(c)} \hfill \vspace{8mm}

{\includegraphics[scale=0.3]{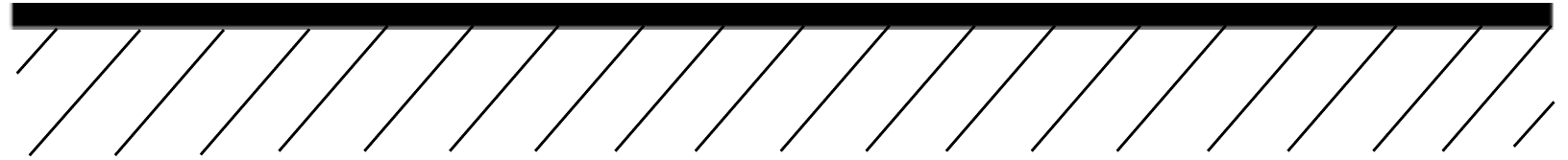}}
\caption{ (color online). A sample surface with radius of curvature $r$ possessing different structures.
        a) A spherical grain of radius $a$ on a sample. The grain's curvature is determined by its radius viz. $k=1/a^2$; b) A bump of radius $a$ on a sample;  
c) Smooth surface.}
\label{fig:imperf}
\end{figure}

Physically, the deviation from the universal behavior occurs because the curvature radii of individual surface structures can be significantly different from the sample's overall radius of curvature. To highlight this effect we consider two elementary examples of surface structures like a single small spherical grain [Figure \ref{fig:imperf}(a)] and a single bump on a flat sample's surface [Figure \ref{fig:imperf}(b)]. 
Surface structures of more complex shape can be reduced to these elementary ones. For comparison we also give the SEE characteristics in the case of a smooth surface [Figure \ref{fig:imperf}(c)].
Here, we deal with the case when the density of surface structures is low, i.e. when the distance between structural elements is much larger than their size. This excludes the effect of re-entrance of emitted electrons into another part of the surface since the probability of this process becomes very small \cite{Nishimura}.

To describe SEE due to isotropic incidence of primary electrons $e^{-}_{p}$ of energy $E_0$ to the objects with varying surface curvature $k$ we generalize a commonly used expression \cite{Jonker1952,Sternglass1957,Chowetal1993} for the electron yield as
\begin{eqnarray}
\label{IsoIncid}
\delta(E_0,k)=\int_0^{R(E_0)}\frac{1}{\epsilon} \left(-\frac{dE}{dx}\right)\Theta\left[\Delta-x\right] dx \times\\
 \nonumber \times \int_0^{\phi_c}{\exp\left({-\frac{l}{\lambda}}\right)\sin \phi \displaystyle d \phi}
\end{eqnarray}
and average over the incidence angle $\theta$
\begin{eqnarray}
\label{IsoAngle}
\left<\delta(E_0,k)\right>_{\theta}=C\int_0^{\theta_c}{\delta(E_0,k)\sin \theta \cos \theta d\theta}.
\end{eqnarray}
Here, following Jonker \cite{Jonker1952}, we consider the case when secondary electrons are generated at a distance $x$ from the entry of the primary electron into a sample and move to the target's surface at an angle $\phi$ with the direction to the nearest surface point. The Heaviside function $\Theta(\Delta-x)$ selects contributions to SEE only from primaries traveling on a straight path lying entirely in the target body. Here $\Delta(k,\theta)$ 
is the linear dimension of the object along the path of the primary, labeled by the direction $\theta$. The surface curvature $k$ is the function of the local curvatures $\chi$ of the object. The parameter $\lambda$ denotes the mean free path of secondary electrons within the target, $\epsilon$ is the energy necessary to produce one secondary electron, $dE/dx$ is the energy loss of the primary per unit path length and $l(x,\phi)$ is the distance that is necessary to reach the sample's surface from the point where secondaries $e^{-}_{s}$ are generated. The normalization constant $C$, the distance $l$, and the limits of integration in Eq. (\ref{IsoAngle}) depend on the object geometry. Equations (\ref{IsoIncid}) and (\ref{IsoAngle}) are very general and can be used to study the SEE from objects of arbitrary shape. In order to be more specific we shall focus on the cases described by [Figures \ref{fig:imperf}(a),(b),(c)]. For all types of surface structures that we consider $C=1$. In equation (\ref{IsoIncid}) for a spherical dust grain [Figure \ref{fig:imperf}(a)] the maximal possible penetration depth for incident direction with angle $\theta$ to the surface normal vector is $\Delta= 2a\cos\theta$ and $R(E_0)=(An)^{-1}E_{0}^{n}$ is the projected range. The parameters $A$ and $n$ depend on the projectile and target and are determined by experimental measurement of $R(E_0)$, giving $n=1.5$ for electrons. In the case of a bump, [Figure \ref{fig:imperf}(b)]  we have $\Delta=a\cos\theta$. In both cases we take a semi-infinite slab as the underlying sample object. Finally, in the case of Figure \ref{fig:imperf}(c), a smooth surface, the maximum depth is always given by $x_m=R(E_0)$. Note that when dealing with a smooth surface of radius of curvature $r$ one can obtain from equation (\ref{IsoIncid}) an expression for the secondary electron yield valid for a semi-infinite slab by taking the limit $x/r\rightarrow0$. 

We emphasize that despite the simplifications introduced by the model of SEE production described by Jonker \cite{Jonker1952}, it does provide a useful approximation to experimentally observed data. It was shown that the combination of this model with a Monte Carlo trajectory simulation allows SEE and BSE yields to be calculated, simultaneously, with good accuracy \cite{Joy}. 

To derive conditions for the applicability of the conventional formulation of SEE we examine how the simplifying assumption of universal dependence is expressed in the framework of the current study. With the help of equation (\ref{IsoAngle}) and using results from \cite{DraineSalpeter1979} we obtain a 
\begin{figure}[ht!]
{\includegraphics[scale=0.3]{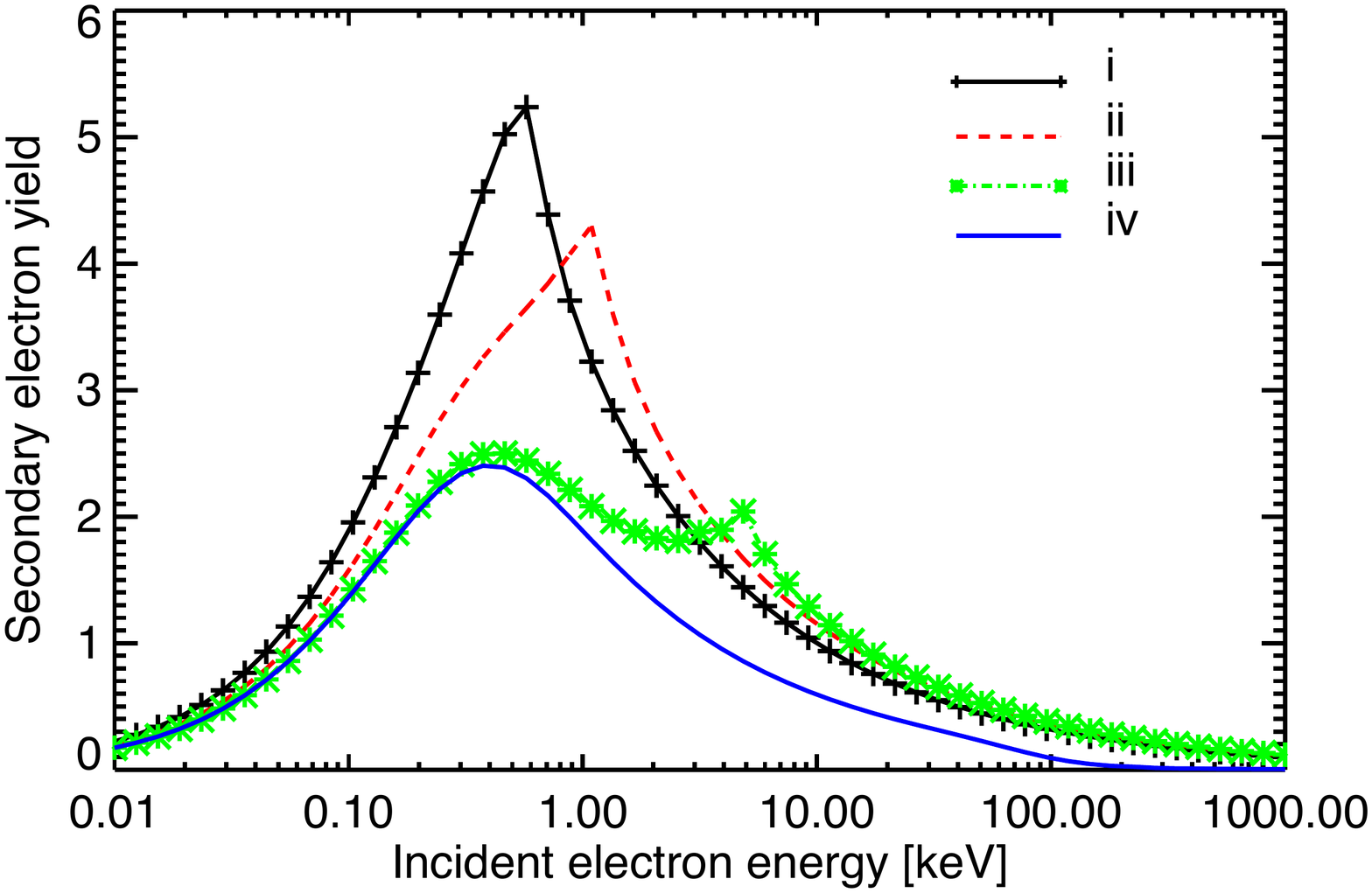}}
{\includegraphics[scale=0.3]{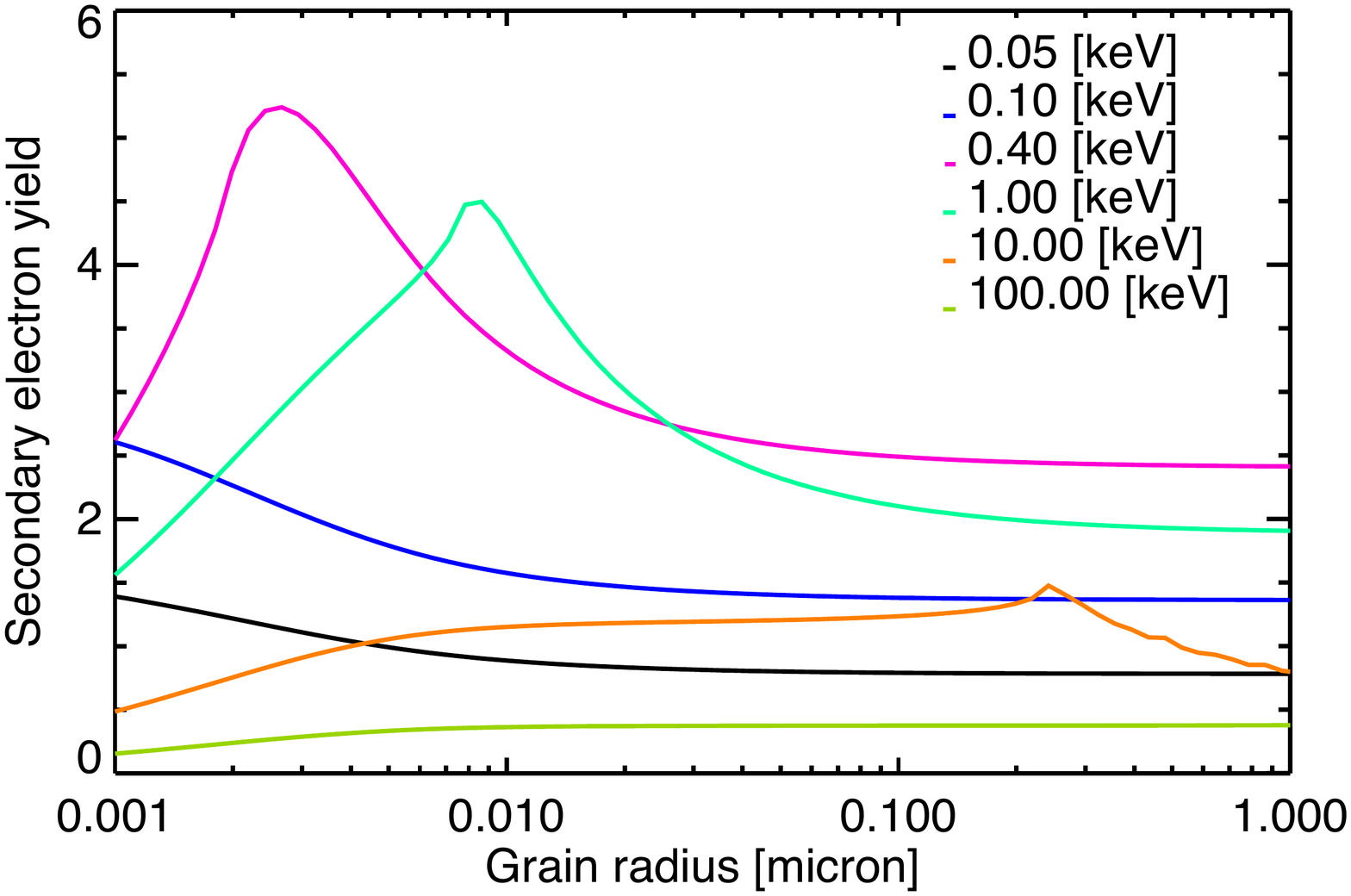}}
\caption{ (color online). (top) SEE yield for isotropic incidence from a silicate grain sitting on a flat substrate (case (a) from Figure \ref{fig:imperf}). Grains of different radii are considered: (i) $a=0.004$ $\mu \text{m}$, (ii) $a=0.01$ $\mu \text{m}$, (iii) $a=0.1$ $\mu \text{m}$ and 
(iv) a silicate grain in the limit of big radius $x/a\rightarrow0$; (bottom) Dependence of SEE yield from silicate grains on the grain radius (case (a) from Figure \ref{fig:imperf}) for various primary electron energies.}
\label{Yields}
\end{figure} 
new generalization of the expression for the SEE yield for a semi-infinite slab
\begin{equation}
\label{YieldMaxIso}
\left<\delta(E_0,k)\right>_{\theta}=\frac{1}{\epsilon}\left(An\lambda\right)^{\frac{1}{n}}\hat{G}_n\left(\left[\frac{R}{\lambda}\right]^{\frac{1}{n}}\right),
\end{equation}
with $\hat{G}_n(\eta)$ given by
\begin{eqnarray}
\label{Giso}
\hat{G}_n(\eta) \equiv\int_0^1{\mu^{1-\frac{1}{n}} d\mu}\int_0^1{\nu^{\frac{1}{n}}exp\left(-\mu\frac{\eta^n}{\nu}\right) d\nu} \times \\
\nonumber \times \int_0^{\mu(\eta/\nu^{\frac{1}{n}})}{exp(\tau^n) d\tau}.
\end{eqnarray}

We find that the function $\hat{G}_n(\eta)$ has a single maximum $\hat{G}_n(\eta_m)$ at $\eta=\eta_m$. 
The mean free path and 
\begin{figure}[ht]
{\includegraphics[scale=0.3]{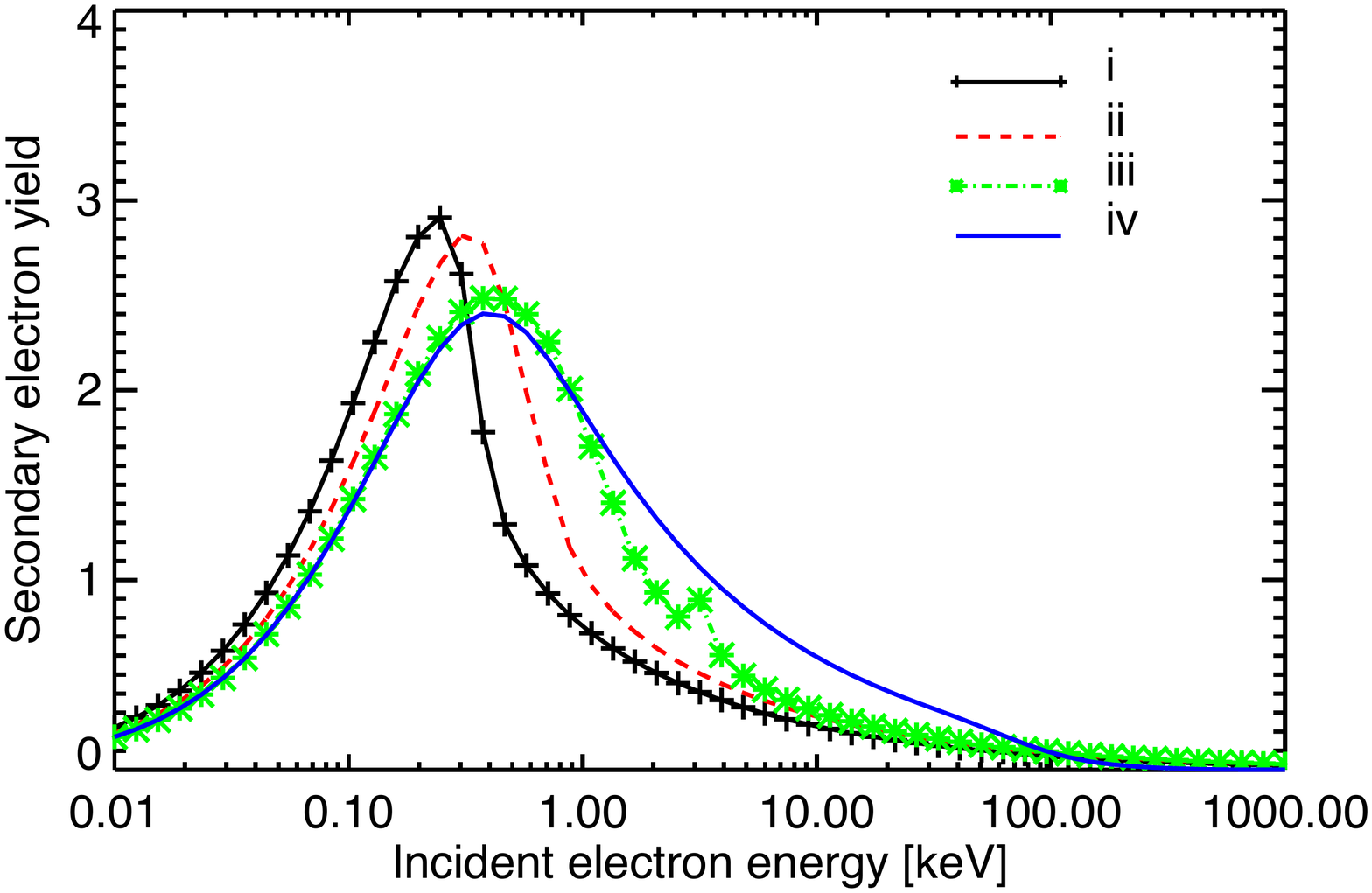}} 
{\includegraphics[scale=0.3]{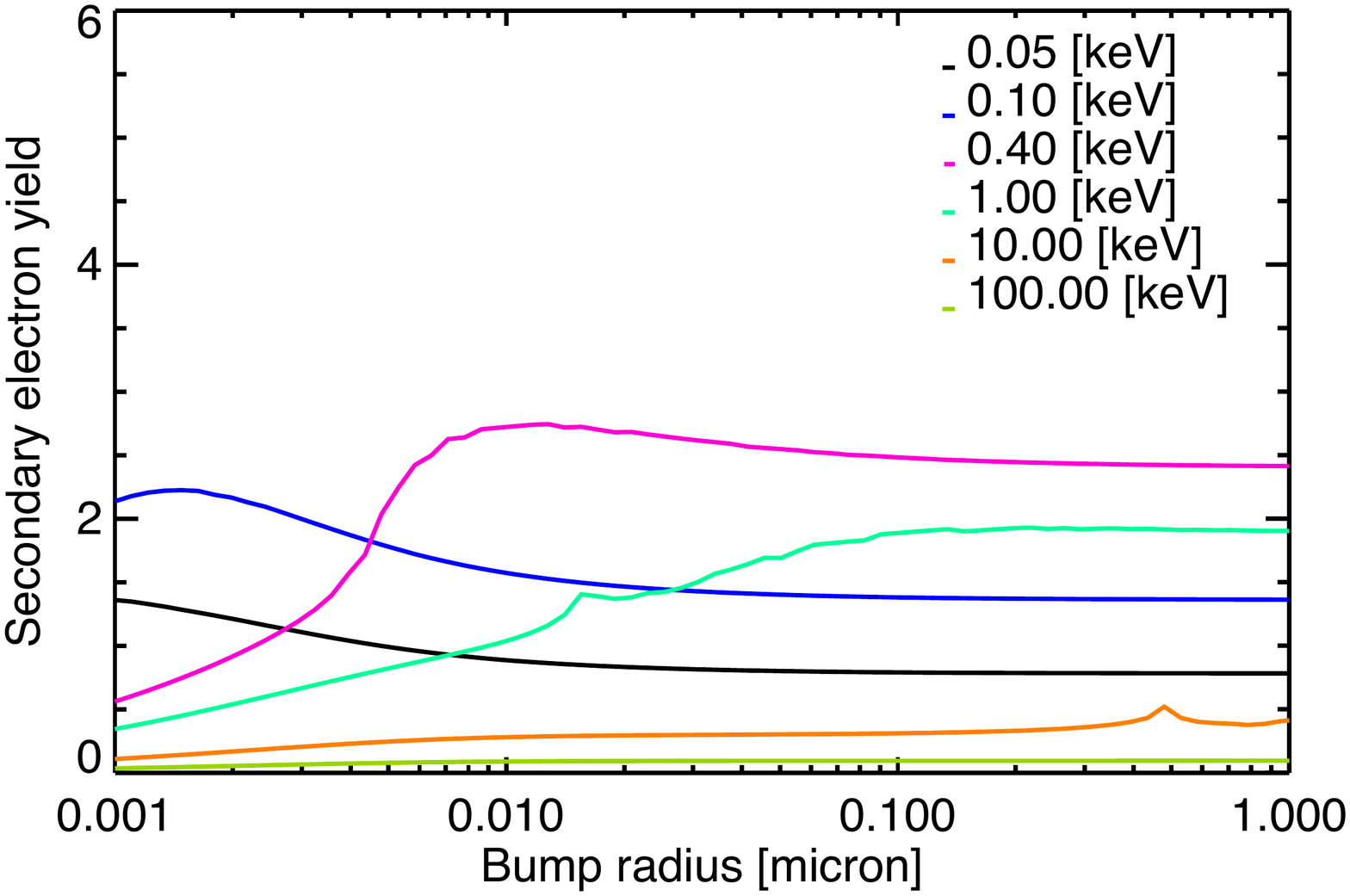}}
\caption{ (color online). 
(top) SEE yield, for mono-directional incidence, from a hemispherical bump sitting on a flat substrate (case (b) of Figure \ref{fig:imperf}).
Bumps of different radii are considered:(i) $a=0.004$ $\mu \text{m}$, (ii) $a=0.01$ $\mu \text{m}$, (iii) $a=0.1$ $\mu \text{m}$ and (iv) a silicate bump in the limit of big radius $x/a\rightarrow0$; (bottom) Dependence of SEE yield from a silicate bump  on the bump radius (case (b) from Figure \ref{fig:imperf}) for various primary electron energies.}
\label{Yields0}
\end{figure}
the dissipation energy can be written as \cite{DraineSalpeter1979}
\begin{equation}
\label{lambdaIso}
\lambda=R(E_m)/\eta_m^n
\end{equation}
and
\begin{equation}
\label{epsilonIso}
\epsilon=(E_m/\delta_m)[\hat{G}_n(\eta_m)/\eta_m],
\end{equation}
where $\delta_m$ is the maximum yield and $E_m$ is the corresponding energy. The parameters $\delta_m$ and $E_m$ can be measured in experiments of SEE from semi-infinite slabs. It is reasonable to assume that the values of mean free path and the dissipation energy are independent on shape and structure of the target. 

\section{Results}
To evaluate equations (\ref{IsoIncid}), (\ref{IsoAngle}), (\ref{lambdaIso}) and (\ref{epsilonIso}) numerically one needs an expression for $R(E_0)$. 
\begin{figure}[ht!]
\centering
\includegraphics[scale=0.3] {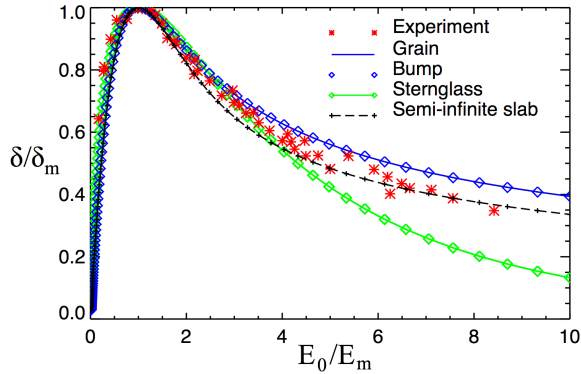}
\caption{ (color online). Secondary electron yield induced by electron bombardment, compared between different SEE models and experiments \cite{Salow, Sternglass1954, Young} for a flat surface. Material parameters for silicate particles are used (Table \ref{Table}).}
\label{Convergent}
\end{figure}
We use the approximation given by Fitting \cite{Fitting}, which reads $R(E_0) = 50 \ \text{nm} \left(\displaystyle\frac{10^3 \text{kg m}^{-3}}{\rho}\right)\left(\displaystyle\frac{E_0}{\text{keV}}\right)^{3/2}$ for material of bulk density $\rho$. This expression is valid over a wide range of electron energies $0.1$ $\text{keV} \lesssim E_0 \lesssim 1$ $\text{MeV}$.  
As an example we consider a silicate. The bulk density for silicate is given by $\rho=3.3 \times 10^3$ $\text{kg}/\text{m}^3$. The parameters $\delta_m$ and $E_m$ for electron bombardment on a semi-infinite slab are taken from experiments, giving $\delta_m=2.4$ at $E_m=400$ $\text{eV}$ for silicate \cite{Kollath}.

To describe the surface sensitivity we use the secondaries escape depth $d(\lambda,\phi)=\lambda \text{cos}\phi$ \cite{Jablonski} -- the distance normal to the surface from which the secondaries escape. Secondary electrons have a very small escape depth (typically $\sim 10 \text{nm}$) due to their low energy. The number of secondaries that migrate to the surface and escape decreases exponentially with depth so that only those produced within a thin surface layer contribute significantly to the observed yield. 
In what follows, we focus on the interplay between primaries penetration depth, the size of the surface structure (surface curvature $k$) and the secondaries escape depth.

Results for the reference cases are illustrated in  Figure \ref{Yields} and Figure \ref{Yields0}.
Figure \ref{Yields} (top) shows the secondary electron yield from a single spherical silicate dust particle that is situated on a flat substrate [Figure \ref{fig:imperf}(a)] and Figure \ref{Yields0} (top) presents the secondary electron yield only from a single hemispherical silicate bump (situated on a flat substrate) [Figure \ref{fig:imperf}(b)] of different radii induced by electron bombardment.
As seen on both figures, a surface structure with curvature radius $a=0.1\mu\text{m}$ and smaller exhibits a yield curve which is generally larger than that of surface structure with larger curvature radii.  In the limit of very large surface structure the SEE yield converges to that from a flat surface, also revealing the universal energy dependence. 
This is consistent with measurements of SEE from polystyrene latex spheres of sub micron size, where the yield was found to agree with the value measured for polystyrene foils \cite{HallBeeman}, \cite{Ziemann}. The highest value for secondary electron yield is obtained when $R(E_m)\approx 1/\chi$, here $1/\chi=a$ is comparable to the maximal escape depth of the  secondaries $d_m=\text{max}\{d(\lambda,\phi)\}|_{\phi}$. This implies that at certain values of the penetration depth $R(E_0)$ the production of secondaries is maximized. In Table \ref{Table} we list some representative SEE characteristics for some typical materials. 

Note that there is a second peak appearing in the yield curves shown in Figure \ref{Yields}(iii) and Figure \ref{Yields0}(iii), which represents an effect due to small surface structures. For $a<1\mu\text{m}$ the projected range at those energies $E_2$ where the additional peak occurs is $R(E_2)\approx2a$ for the grain and $R(E_2)\approx a$ for the bump. Thus, peak forms, when at some $\theta$ the linear dimension $\Delta(k,\theta)$ of the surface structure along the path of the primary is maximized and the projected range: $R(E_2)\approx \text{max}\{\Delta(k,\theta)\}|_{\theta}\gtrsim d_m$. 
The peak appears as long as $R(E_m) \lesssim \text{max}\{\Delta(k,\theta)\}|_{\theta}$. A similar second peak at $E_0 > E_m$ was also found in the yield curve of the SEE from surfaces of carbon foils \cite{Caron}. 

Figure \ref{Yields} (bottom) and Figure \ref{Yields0} (bottom) show the effect of the primary electron energy on the SEE yield depending on the size of the surface structure. The effect of small surface structures becomes more pronounced as the electron energy increases from approximately $100\text{ eV}$ to few $\text{keV}$. The contribution of the curvature to the size dependence of the yield therefore dominates at intermediate electron energies.
For small ($a<0.1\mu\text{m}$) surface structures the SEE yield does not reveal the universal dependence on energy as is observed for large ($a>1\mu\text{m}$) surface structures and flat surfaces. 
Figure \ref{Convergent} illustrates this dependence for the surface types shown in Figure \ref{fig:imperf} (adhered spheres, half-spheres, and a flat surface) for a silicate material (parameters from Table I). Experimental data for a flat silicate surface are also shown.
\begin{table}[htbp]
\begin{center}
\caption{SEE parameters for different materials} \label{Table}
\vspace*{.5cm} 
\begin{tabular}{l*{6}{c}r}
\hline 
\vspace*{.05cm} Model &&\multicolumn{1}{c}{{\rotatebox[origin=c]{270}{ \hspace*{.4cm}Material}}}& $E_m$   &   $\delta_m$   &   $R(E_m)$\\
\vspace{.05cm}&&&(eV)&&($\mu\text{m}$)& \vspace{.1cm}\\
\hline 
\\ 
Kollath, 1956 \cite{Kollath}&&silicate\footnote{$\text{SiO}_2$}&  400 & 2.4 & 0.0038  \vspace{.05cm}\\
Hachenberg,  Brauer, 1959 \cite{Hachenberg}          &  & Ca&300 & 1.5 & 0.0028 \vspace{.05cm}\\
Draine, Salpeter, 1979 \cite{DraineSalpeter1979}           & & Ice& 500 & 2.0 & 0.019   \\
\\
\hline 
\end{tabular}
\end{center}
\end{table}

As a result, SEE from a spherical grain of given curvature with isotropic incident flux converges in the limit of big grain 
radius to the case of SEE from a smooth surface, a semi-infinite slab model. This convergence of $\left<\delta^{grain}\left(E_0,k=1/a^2\right)\right>_{\theta}$ to $\delta^{slab}(E_0,k=0,\theta=0)$ 
can be seen in Figure \ref{Convergent}. The same result is obtained for SEE from a hemispherical bump with mono-directional incident flux 
[Figure \ref{Convergent}]. 

\section{Conclusions}
Generally, the values of the secondary electron yield from large surface structures with $a > 1\mu$$\text{m}$ show no further variation with the size of surface structure and are identical to those from a semi-infinite slab model. In contrast, surface structures with curvature radii from nanometers to sub-microns may have a strong effect on SEE. It is also reasonable to mention that the SEE yield from the surface with distributed structures of low concentration is proportional to the concentration of the structures \cite{Branko2008,Branko2012} and greater than the SEE yield from the smooth surface as long as $0.01\lesssim a/L\lesssim1$, here $L$ is a distance between structural elements \cite{Nishimura}.

{\bf Acknowledgments}
This work was supported by Deutsche Forschungsgemeinschaft through Schwerpunktprogramm 1488 "Planet Mag" (Grants SCHM 1642/2-1 and 
YA 349/1-1).

\end{document}